\documentclass[intlimits,twoside,a4paper]{article}  

\usepackage{amsmath,amssymb}
\usepackage{graphicx}
\usepackage{siunitx}

\usepackage[T2A]{fontenc}
\usepackage[cp1251]{inputenc}

\usepackage[eqsecnum]{cmpj2}

\articletype{Proceedings Paper}

\issue{2013}{16}{3}{31801}
\doinumber{10.5488/CMP.16.31801}

\title[Chemical capacitance for manganite based ceramics]{
Chemical capacitance proposed for \\ manganite-based ceramics
}

\author[A.~Molak]{A.~Molak}
\address{
Institute of Physics, University of Silesia, 4 Uniwersytecka St., 40--007~Katowice, Poland
}

\date{Received November 7, 2012}
\authorcopyright{A.~Molak, 2013}

\begin{document}

\maketitle

\begin{abstract}
The measured value of effective electric permittivity $\varepsilon_\mathrm{eff}$  of several compounds, e.g., (BiNa)(MnNb)O$_3$, \linebreak (BiPb)(MnNb)O$_3$, and BiMnO$_3$ increases from a value $\approx 10\div100$ at the low temperature range ($100\div300$~K) up to the high value reaching the value $10^5$ at high temperature range, e.g., $500\div800$~K. Such features suggest the manifestation of thermally activated space charge carriers, which effect the measured capacitance. The measured high-value effective permittivity of  several manganite compounds can be ascribed to the chemical capacitance
 $C_{\mu} = e^2 \partial N_{i}/\partial\mu_i $
expressed in terms of the chemical potential $\mu$.
The chemical capacitance $C_{\mu}^{(\textrm{cb})} = e^2 n_\mathrm{C}/k_\mathrm{B}T$ depends on
temperature when the conduction electrons with density $ n_\mathrm{C} = N_\mathrm{C} \exp\left(\mu_{n}- E_\mathrm{C}\right)/k_\mathrm{B}T$
are considered. The experimental results obtained for the manganite compounds, at high temperature range, are discussed in the framework of the chemical capacitance model. However, the measured capacitance dependence on geometrical factors is analysed for BiMnO$_3$ indicating that the non-homogeneous electrostatic capacitor model is valid in $300\div500$~K range.
\pacs  82.45.Un,  77.22.Ch
\end{abstract}

\section{Introduction}

Recently, scientists have turned their attention to the development of high power electrical energy storage devices. Such a system includes electrochemical and electrostatic capacitors. The promising materials are oxides and composites of oxides, e.g., MnO$_2$, RuO$_2$, Fe$_2$O$_3$ \cite{1-Sherill2011}. The occurrence of defects produces an effect on the electric properties of a sample. Primarily, they affect the electronic structure and thus the conductivity. However, due to the relation $\varepsilon^* = \ri \sigma^*$, the effective value of electric permittivity is usually measured in case of the materials with perovskite structure that contain defects. In such a case, contribution of the lattice and the space charge or charge carrier responses ought to be considered \cite{2-Molak2009}
\begin{equation}
\varepsilon^* (\omega) = \varepsilon^*(\omega)_\textrm{lattice} + \varepsilon^*(\omega)_\textrm{carries}\, .
\label{e1}
\end{equation}

The concept of a capacitance is usually related to an electrostatic geometric capacitor determined by the electric field $E$ between two metal electrodes storing opposite charges. When a dielectric fills the capacitor, the electric field is modified due to the appearance of induced and ordered dipoles. Hence, the capacitance of such a system can be modified in accord with the dielectric material properties
\begin{equation}
C_\textrm{electrostatic} = \varepsilon_ r\varepsilon_0 \frac{A}{d}\, .
\label{e2}
\end{equation}
However, the excess charge is confined to the thin region close to the interface between dielectric material and the electrode, and the sub-electrode layer can be considered. The contribution of defects can be considered basing on a multi-layer condenser \cite{3-Mitsui1976}. In such a case, the space charge effects are ascribed to the sub-electrode layers whose properties are different from those of the bulk.

The perovskite stoichiometric ABO$_3$ compounds, e.g., niobates, tantalates and titanates, show a large gap $E_\mathrm{g}  \sim3$~eV in the electronic structure. Hence, their dielectric properties are described using the electrostatic geometric capacitance [equation (\ref{e2})]. However, several perovskites that contain the $3d$ transition metal ions in the B sublattice, e.g., BiMnO$_3$, (BiNa)(MnNb)O$_3$ \cite{4-Molak2005,5-Molak2006}, (BiPb)(MnNb)O$_3$ \cite{6-Molak2005}, have a narrow gap $E_\mathrm{g}<1$~eV.
Therefore, they show a marked conductivity at temperature ranges above the room temperature. It was reported that these perovskite materials (see also references in \cite{4-Molak2005,5-Molak2006,6-Molak2005} for other materials) exhibit the effective permittivity which reaches high values $\varepsilon_\mathrm{eff}\sim 10^{5}\div10^6$ at high temperature when measured at radio-frequencies. This effect can be ascribed to the occurrence of defects, not only to electric current charges and to oxygen vacancies but also to chemical and structural non-homogeneity. Such features correspond to the semiconductor behaviour of the electric conduction. The reported activation energy values varied within  $0.2\div 1.0$~eV range \cite{4-Molak2005,5-Molak2006,6-Molak2005}.

It seems worthwhile to discuss whether the electric properties of the non-homogeneous perovskite compounds can be considered in terms of the chemical capacitance. Such an approach would be limited, e.g., to the manganite-based compounds, their solid solutions, and to compounds containing a high amount of oxygen vacancies.

The electrochemical capacitance includes electrical and chemical contributions
\begin{equation}
C_{\textrm{electrochem}}   \, {\varpropto} \,  \left(\frac{\partial {\phi}}{\partial Q} + \frac{\partial {\mu}^{*}}{\partial Q}\right)^{-1}
\,  {\varpropto} \, \left(\frac{1}{C_{\textrm{electro}} } +  \frac{1}{C_{\textrm{chem}} }\right)^{-1},		
\label{e3}
\end{equation}
where $Q$ denotes the charge, $\phi$~--- the electric potential at the electrode, $\mu^{*} = (1/ez)\mu$ denotes the normalized chemical potential, $e$~--- the elementary charge, and $z$~--- charge number (the notation which discriminates the type of the charge carriers and non-homogeneity of the material is omitted here for simplicity) \cite{7-Jamnik2001}.

The general electrochemical capacitance has been defined for mesoscopic systems \cite{8-Bisquert2003}. The capacitance relates to the density of states
\begin{equation}
C_{\textrm{DOS}} = e^{2}\frac{\rd N}{\rd E}\, ,
\label{e4}
\end{equation}
where $N$ denotes the density of charges. The other definition of the chemical capacitance is related to the carriers density $c$, since $\mu = \mu^{*} + k_\mathrm{B} T \ln(c/N)$,
\begin{equation}
C_\textrm{chem} = (ez)^{2} \left(\frac{\partial {\mu}}{\partial c}\right)^{-1} Ad =  \left[ \frac{(ez)^{2}}{k_\mathrm{B}T} \right] c Ad\, .
\label{e5}
\end{equation}	
Therefore, the chemical capacitance is proportional to the volume, $V = Ad$, which contains the charge carriers.
Correspondingly, the electrochemical resistor can be defined and the electric current density was obtained \cite{7-Jamnik2001}
\begin{equation}
I(r,t) =  z e J  = -\sigma (r) \nabla\left(\mu^{*}  + \phi\right).
\label{e6}
\end{equation}

This idea of the chemical capacitance was proposed e.g., for the solar cells based on TiO$_2$ nanocomposites working on the redox processes \cite{8-Bisquert2003, 9-Fabregat2011} and (La,Sr)(Co,Fe)O$_{3-\delta}$ electrodes applied for solid oxide fuel cells \cite{10-Baumann2006}. In such cases, the modification of the electrochemical potential $\rd V$ produces a change in the chemical potential $\rd\mu_{n} $
of electrons, associated both with the variation of free electron density $\rd n_\mathrm{C}$
 and the variation of a localized electron density in band gap states $\rd n_\mathrm{L}$. The chemical capacitance defined locally for a small volume element, reflects the capability of a system to accept or release additional charge carriers with the density $N_{i}$ due to a change in their chemical potential $\mu_{i} = \mu_{i}^{0} + k_\mathrm{B}T$. The chemical capacitance per unit volume is formulated as follows:
\begin{equation}
C_{\mu}^{(i)} =  e^2 \frac{\partial{N_i}}{\partial \mu_i }= \frac{e^2 N_{i}}{ k_\mathrm{B}T}\, .
\label{e7}
\end{equation}
Hence, when both free and localized electrons are considered, one obtains the chemical capacitance dependent on temperature
\begin{equation}
C_\mu = e^2 \frac{\partial (n_\mathrm{C} + n_\mathrm{L} )}{\partial \mu_n}  = C_{\mu}^{(\textrm{cb})} + C_{\mu}^{(\textrm{trap})}\, ,
\label{e8}
\end{equation}	
 where
\begin{equation}
C_{\mu}^{(\textrm{trap})}  = e^2 \frac{\partial{n_\mathrm{L}}}{\partial \mu_n } = e^2 g(\mu _n)\, .
\label{e9}
\end{equation}	
The density of localized states in the band gap, at energy $E$ and for exponential distribution is \cite{8-Bisquert2003, 11-Bisquert2008, 12-Bisquert2008}
\begin{equation}
 g(E) = \left(\frac{N_\mathrm{L}}{k_\mathrm{B} T_{0}}\right) \exp\left(\frac{E - E_\mathrm{C}}{k_\mathrm{B}T_0} \right)\, ,
\label{e10}
\end{equation}
where $N_\mathrm{L}$ is the total density below the conductivity band and $T_{0}$ is a parameter with temperature units that determines the depth of distribution.
Similarly, for electrons placed in a conduction band
\begin{equation}
C_{\mu}^{(\textrm{cb})} = \frac{e^2 n_\mathrm{C}}{k_B T}\, ,
\label{e11}
\end{equation}	
where in accord with the Boltzmann distribution function
\begin{equation}
n_\mathrm{C} = N_\mathrm{C} \exp\left[\frac{\mu _n - E_\mathrm{C}}{k_B T}\right]\, . 	
\label{e12}
\end{equation}	
Therefore, the occurrence of the free charge carriers offers a thermally activated contribution to the chemical capacitance.

\section{Experimental}

Several compounds were chosen for analysis, i.e., BiMnO$_3$, (BiNa)(MnNb)O$_3$, (BiPb)(MnNb)O$_3$, and NaNbO$_3$. The ceramic samples of the (BiNa)(MnNb)O$_3$ and (BiPb)(MnNb)O$_3$ compounds were prepared by high-temperature sintering. The details have been described in the former papers \cite{4-Molak2005,5-Molak2006,6-Molak2005}. Standard electrical properties, e.g., electric permittivity,  dielectric loss coefficient, electric conductivity temperature and frequency dependencies of the (BiNa)(MnNb)O$_3$, (BiPb)(MnNb)O$_3$ compounds have been already published. High values of the effective permittivity were observed for a series of BiMnO$_3$-NaNbO$_3$ compounds. Therefore, the analysis of the end members of this series, i.e., BiMnO$_3$ and NaNbO$_3$, was conducted as well.

The geometrical dependence of the capacitance was measured for BiMnO$_3$ and NaNbO$_3$. Several ceramics BiMnO$_3$ samples with different thickness $d$ and surface area $A$ were prepared. Moreover, one chosen sample of BiMnO$_3$ ceramics was polished step by step to obtain a series with $d$ varying from 3.7~mm to 0.833~mm while the surface $A = 0.72$~mm was constant. The capacitance $C$ and conductance $G$ of the BiMnO$_3$ samples were measured using a HP 4263B LCR meter in a parallel mode.

The results obtained for the NaNbO$_3$ and NaNbO$_{3-x}$ crystals were analyzed for comparison. The crystals have been obtained from the solution of melted salts \cite{13-Badurski1979}. The amount of the oxygen vacancies was increased by the electrochemical procedure conducted at high temperature (950~K). The samples were reduced at a chamber in the air at a pressure lowered to 0.1~Pa. The concentration of the space charge was estimated from depolarization currents. The electric measurements of sodium niobate crystals were conducted using a Tesla BM 595 capacitance meter at $f = 1$~MHz.

\section{Results}

\subsection{Capacitance dependence on temperature}

The former studies conducted for the (BiNa)(MnNb)O$_3$ and (BiPb)(MnNb)O$_3$ compounds showed their chemical and structural disorder. The XRD test showed the coexistence of long range electric order and short-range disorder related to defects \cite{14-Molak2008}. The microanalysis carried out with the SEM tests showed a chemical non-homogeneity in the micro-scale \cite{4-Molak2005,5-Molak2006,6-Molak2005,15-Spolnik2005}. The analysis of the electric conductivity and the effective permittivity was consistent with these features of the materials. The electric conductivity temperature dependences were described within the small polaron model with the activation energy varying in the $0.2\div 0.5$~eV range. The model of a small variable hopping polaron, which has been proposed for these compounds, indicated that a distribution of traps with different energies takes place in the (BiNa)(MnNb)O$_3$ and (BiPb)(MnNb)O$_3$ ceramics.

The effective permittivity $\varepsilon_\mathrm{eff}(f,T)$  estimated from the geometrical formula (\ref{e2}) showed a steep increase in the high temperature range, above 500~K. Moreover, a marked dispersion in $\varepsilon_\mathrm{eff} $  took place. The permittivity estimated at the measuring frequency $f = 100$~kHz, showed moderate values between $\varepsilon_\mathrm{eff} \backsimeq 100\div1000$. On the other hand, the permittivity at the frequency of 100~Hz reached high values about $\varepsilon_\mathrm{eff} \backsimeq 10^{5} \div 10^6$ in the high temperature range $600 \div 800$~K. This effect was ascribed to the mutual effect of the conductivity and the capacitance in accord with the general dependence $\varepsilon ^* = \ri \sigma ^*$.

However, another approach, based on the chemical capacitance concept is considered herein. This model includes the participation of defects and electric current carriers  \cite{7-Jamnik2001, 8-Bisquert2003, 9-Fabregat2011, 10-Baumann2006, 11-Bisquert2008,  12-Bisquert2008}. The occurrence of the point defects, which form the traps, consecutively enables the participation of thermally activated charge carriers in the measured capacitance of the samples.

\begin{figure}[!b]
\centerline{
\includegraphics[width=0.5\textwidth]{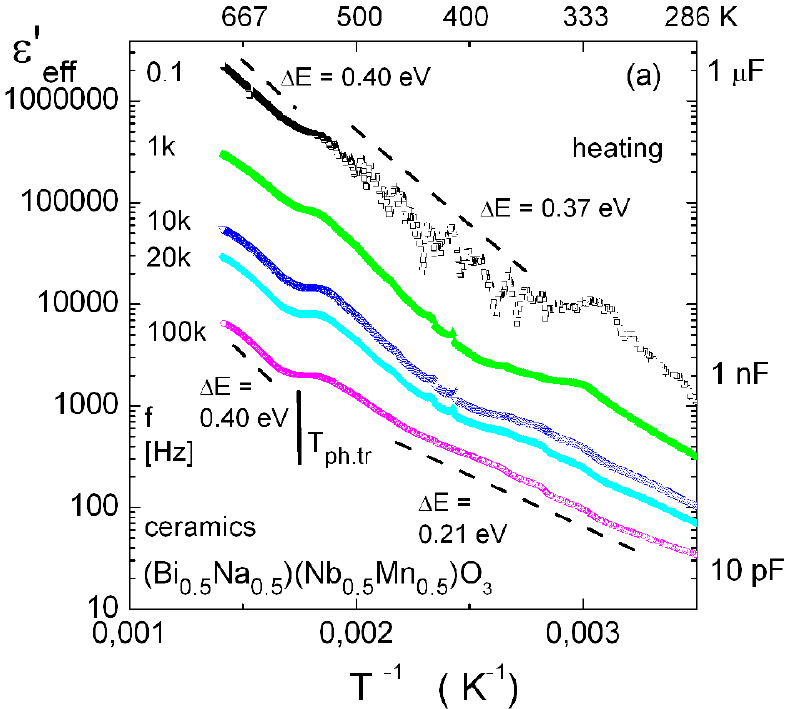}
\hspace{2mm}
\includegraphics[width=0.47\textwidth]{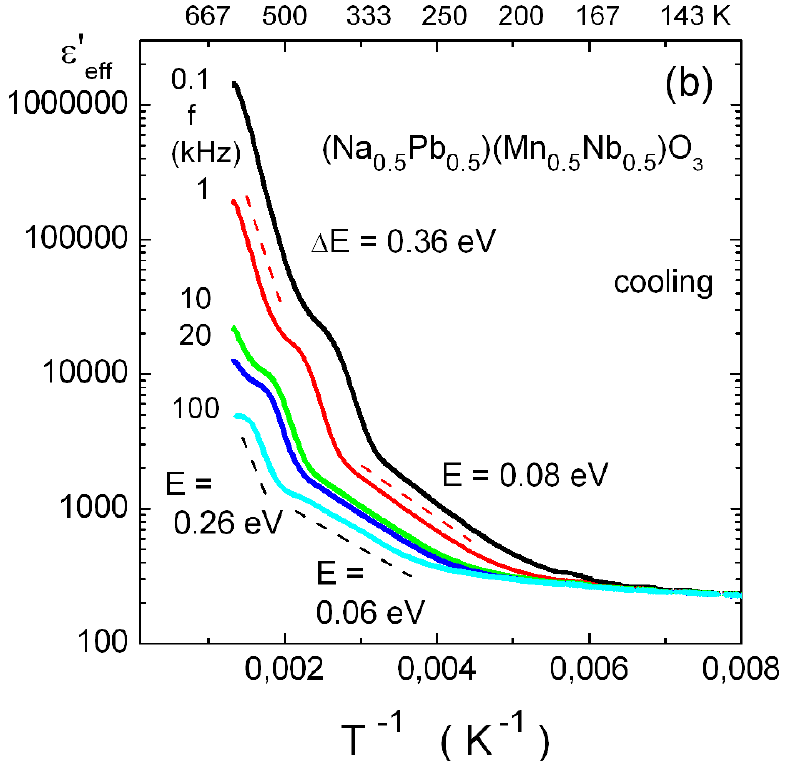}
}
\caption{(Color online) The Arrhenius plots of the effective electric permittivity $\varepsilon '_\mathrm{eff}$ drawn in accord with the chemical capacitance model obtained for (Bi$_{0.5}$Na$_{0.5}$)(Mn$_{0.5}$Nb$_{0.5}$)O$_3$ (a) and (Na$_{0.5}$Pb$_{0.5}$)(Mn$_{0.5}$Nb$_{0.5}$)O$_3$ (b) ceramics.
The parameter $\Delta E$ value and the straight-line segment denote the proposed applicability of the chemical capacitance model.}
\label{fig1}
\end{figure}

\begin{figure}[!t]
\centerline{
\hspace{1mm}
\includegraphics[width=0.503\textwidth]{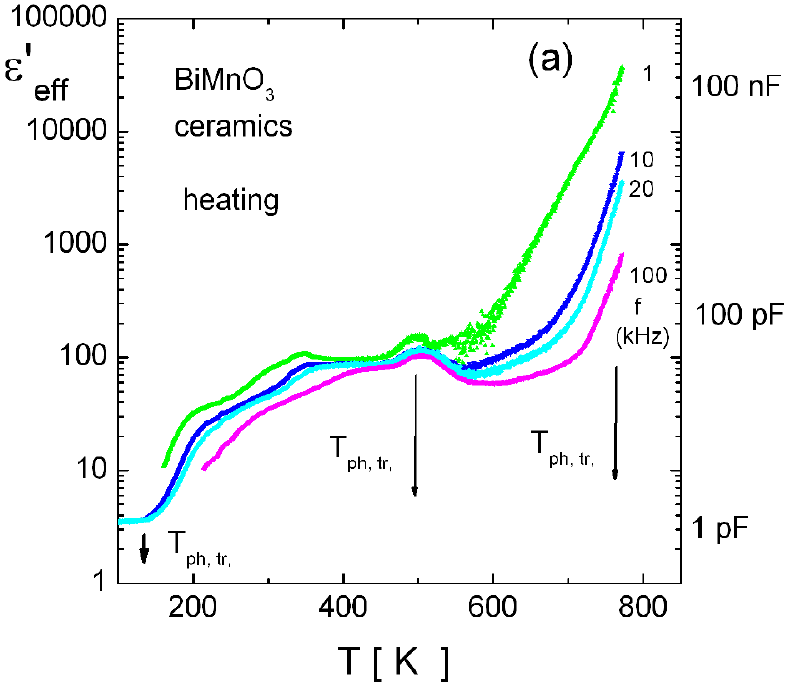}
\hfill
\includegraphics[width=0.47\textwidth]{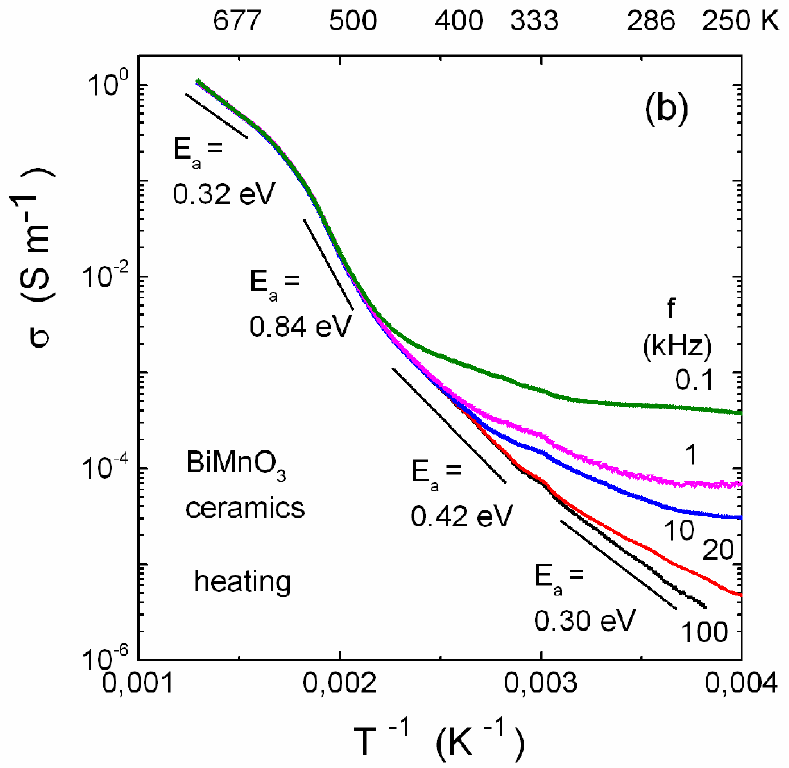}
}
\vspace{1mm}
\centerline{
\includegraphics[width=0.475\textwidth]{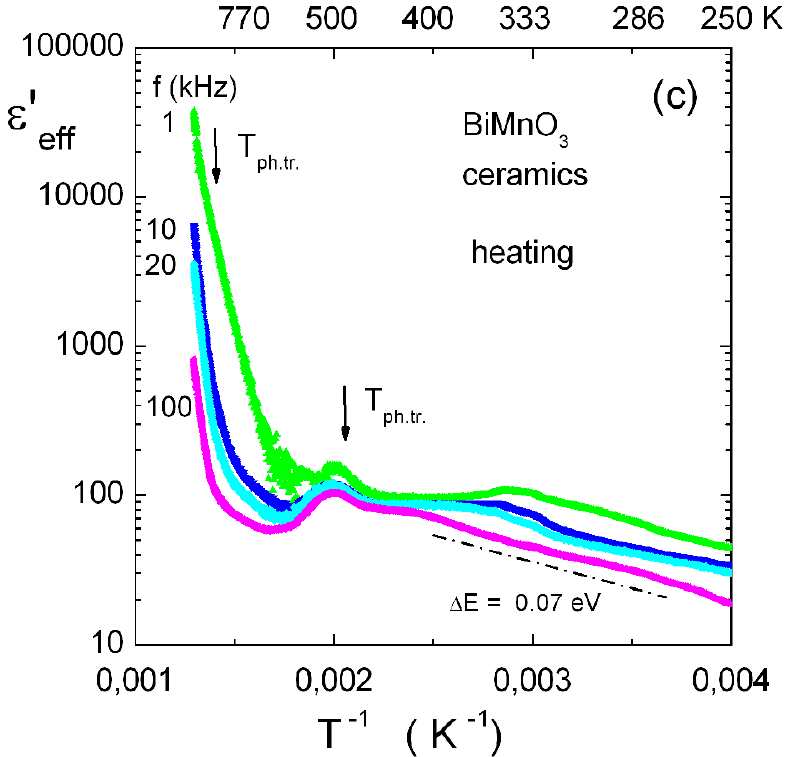}
\hfill
\includegraphics[width=0.45\textwidth]{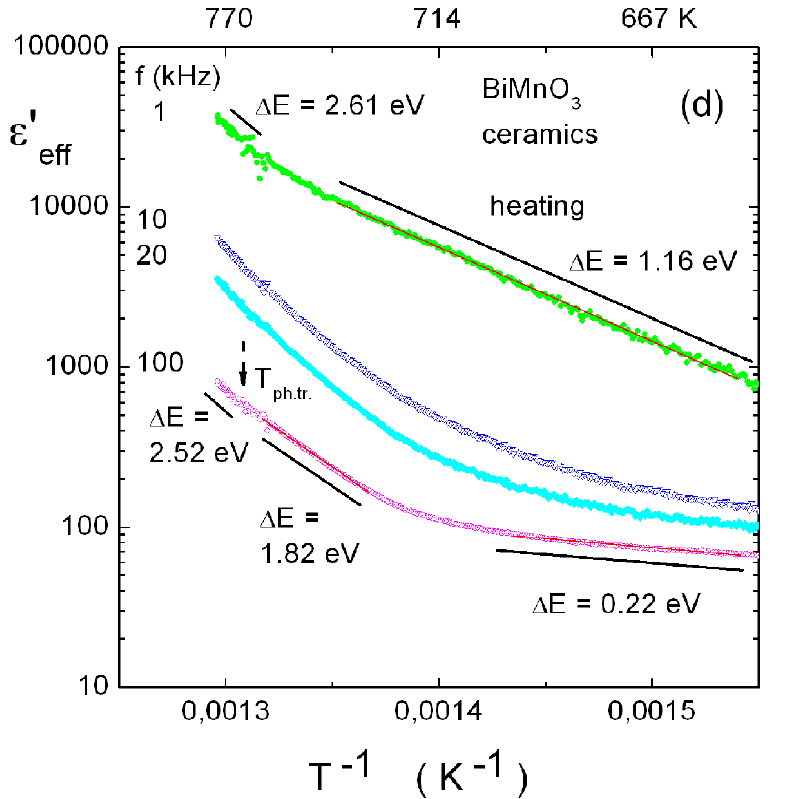}
\hspace{1mm}
}
\caption{(Color online) The plots obtained for BiMnO$_3$ ceramics. (a) The effective electric permittivity temperature dependence $\varepsilon '_\mathrm{eff}$ vs. \textit{T}. (b) The Arrhenius plot of the electric conductivity $\sigma$ vs. $T^{-1}$. (c) The Arrhenius plot of the effective electric permittivity $\varepsilon '_\mathrm{eff}$ obtained for a wide temperature range $250\div800$~K. (d) The Arrhenius plot of effective electric permittivity $\varepsilon '_\mathrm{eff}$ shown in the high temperature range.
\label{fig2}}
\end{figure}

\looseness=-1The Arrhenius-type dependences of the measured capacitance $C$ on temperature $T$ were plotted to check the contribution of the defect subsystem (figure~\ref{fig1} and \ref{fig2}). It turned out that the $C(T,f)$ dependences can be described in accord with the chemical capacitance model  [equations (\ref{e7}), (\ref{e8}), (\ref{e9}),  (\ref{e10}), (\ref{e11}), (\ref{e12})] within the high temperature range. The local peaks or anomalies caused by a phase transition are also visible, e.g., the structural phase transition at $T_\textrm{ph.tr.}= 474$~K manifests itself in case of BiMnO$_3$~\cite{16-Belik2007, 17-Kimura2003}.

The parameter $\Delta E = \left(\mu_n -E_\mathrm{C}\right) $ would be estimated, using a numerical fitting in accord with $n_\mathrm{C} = N_\mathrm{C} \exp[(\mu _n -E_\mathrm{C} )/k_\mathrm{B}T]$ [equation (\ref{e12})], from the slope of straight-line segments discerned in the $\varepsilon '_\mathrm{eff}$ vs. $T^{-1}$ plot. The $\Delta E$  value varies from 0.06~eV in the temperature range in the vicinity of room temperature up to 2.61~eV  in the $765\div 773$~K range depending on the compound (see figures~\ref{fig1} and \ref{fig2}, table~\ref{Table1}). One can notice that the effective permittivity can be described in such cases within the chemical capacitance $C_\mu ^{(\textrm{cb})}$  model
[equations (\ref{e11}), (\ref{e12})] \cite{7-Jamnik2001, 8-Bisquert2003, 9-Fabregat2011, 10-Baumann2006, 11-Bisquert2008,  12-Bisquert2008}.

%

\subsection{The effect of the concentration of oxygen vacancies on capacitance}

\begin{figure}[!b]
\centerline{
\includegraphics[width=0.45\textwidth]{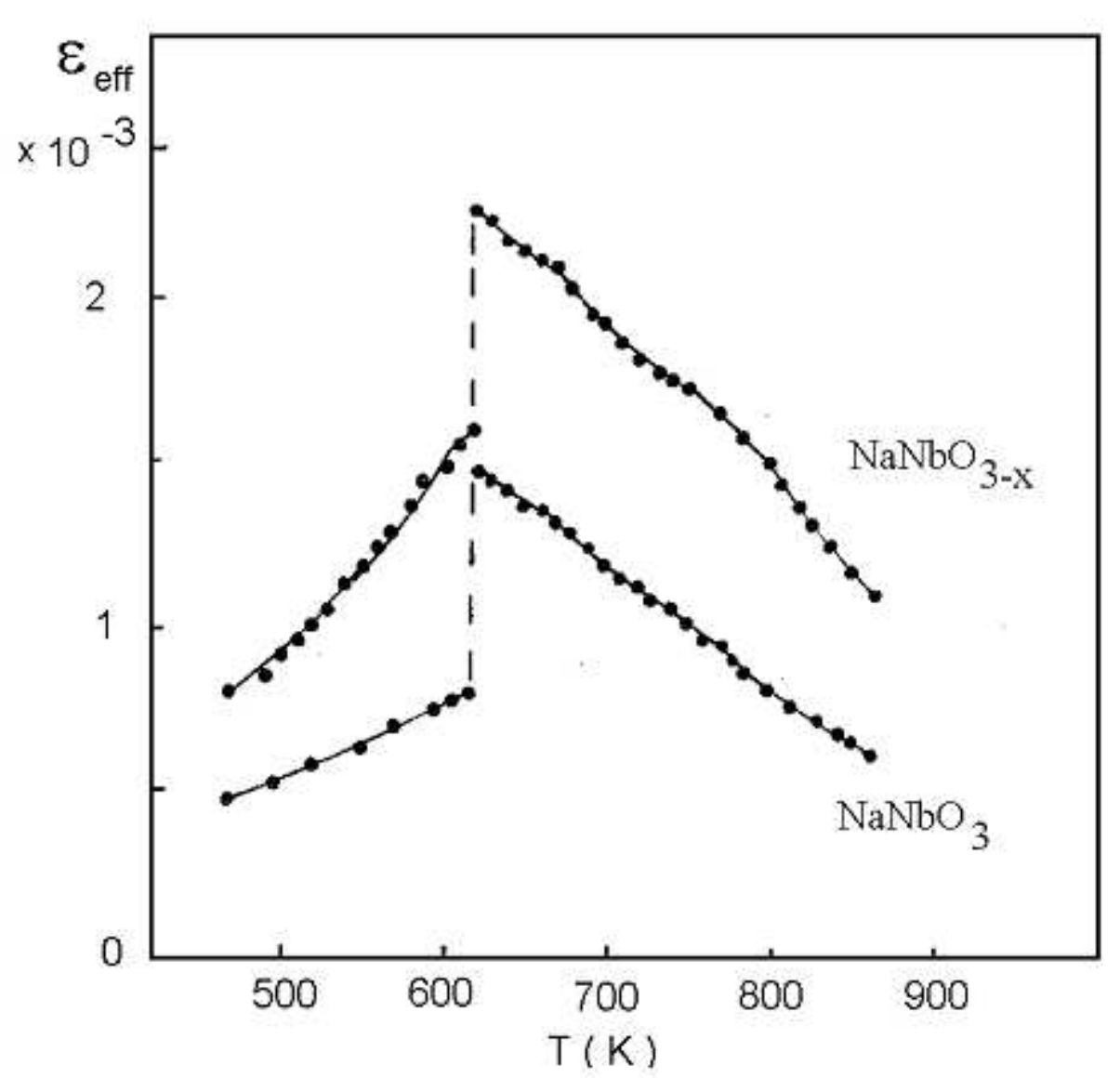}
}
\caption{The electric permittivity $\varepsilon '_\mathrm{eff}$ dependence on temperature obtained for the as-grown NaNbO$_3$ and the oxygen vacancies containing reduced NaNbO$_{3-x}$ crystals.}
\label{fig3}
\end{figure}

The effect of the concentration of oxygen vacancies  on the capacitance is shown for the case of sodium niobate crystals. The estimated concentration of the space charge was equal to $N_{Q} \in (1\cdot 10^{17} \div 6 \cdot 10^{17})$~cm$^{-3}$ in case of the as-grown crystals. The estimated concentration of the charge carriers was one order higher, and it can be ascribed to the appearance of the oxygen vacancies $N_{Q} \backsimeq N_{(\textrm{VO})} \in (2 \cdot 10^{17} \div 1\cdot 10^{18})$~cm$^{-3}$. The $\varepsilon{'}_\mathrm{eff}(T)$ plots obtained at frequency $f = 1$~MHz are shown in figure~\ref{fig3}. A higher amount of the oxygen vacancies resulted in an increase of the actual capacitance, and thus in an increase of the estimated effective permittivity $\varepsilon{'}_\mathrm{eff}$. This effect corresponded to a similar increase of the \emph{ac} conductivity $\sigma_\textrm{ac}(T)$.

\begin{table}[!t]
\caption{Data obtained for the ceramics compounds. The parameter $\Delta E$ value estimated in the chosen temperature ranges $\Delta T$ from numerical fit in accord with the chemical capacitance model. The $\varepsilon '_\mathrm{eff}$ vs. reciprocal temperature $T^{-1}$ plots are shown in figures~\ref{fig1} and \ref{fig2}.}
\label{Table1}
\vspace{2ex}
\centering
\begin{tabular}{|c|c|c|c|c|}\hline
Compound & \multicolumn{2}{|c|}{$f = 100$ kHz} & \multicolumn{2}{|c|}{$f = 1$ kHz} \\
\cline{2-3} \cline{4-5} 
& $\Delta E (\text{eV}) $ & $\Delta T (\text{K})$ & $\Delta E (\text{eV}) $ & $\Delta T (\text{K})$\\  \hline\hline
BiMnO$_3$         & 2.52 & 765--773  & 2.61 &765--773 \\
		& 1.82 & 730--760 & 1.16 & 650--740 \\
		& 0.22 & 650--670 &  &  \\
		& 0.07 & 260--400 &  & \\
(Bi$_{0.5}$Na$_{0.5}$)(Mn$_{0.5}$Nb$_{0.5}$)O$_3$	 & 0.40 & 600--690 & 0.40 & 600--710 \\
		               & 0.21 & 310--400 & 0.37 & 360--520 \\
(Na$_{0.5}$Pb$_{0.5}$)(Mn$_{0.5}$Nb$_{0.5}$)O$_3$ & 0.26 & 530--690 & 0.36 &530--690 \\
			& 0.06 & 270--430 & 0.08 & 220--340 \\  \hline
\end{tabular}
\end{table}

\subsection{The effect of Mn addition on capacitance}

\begin{figure}[!t]
\centerline{
\includegraphics[width=0.9\textwidth]{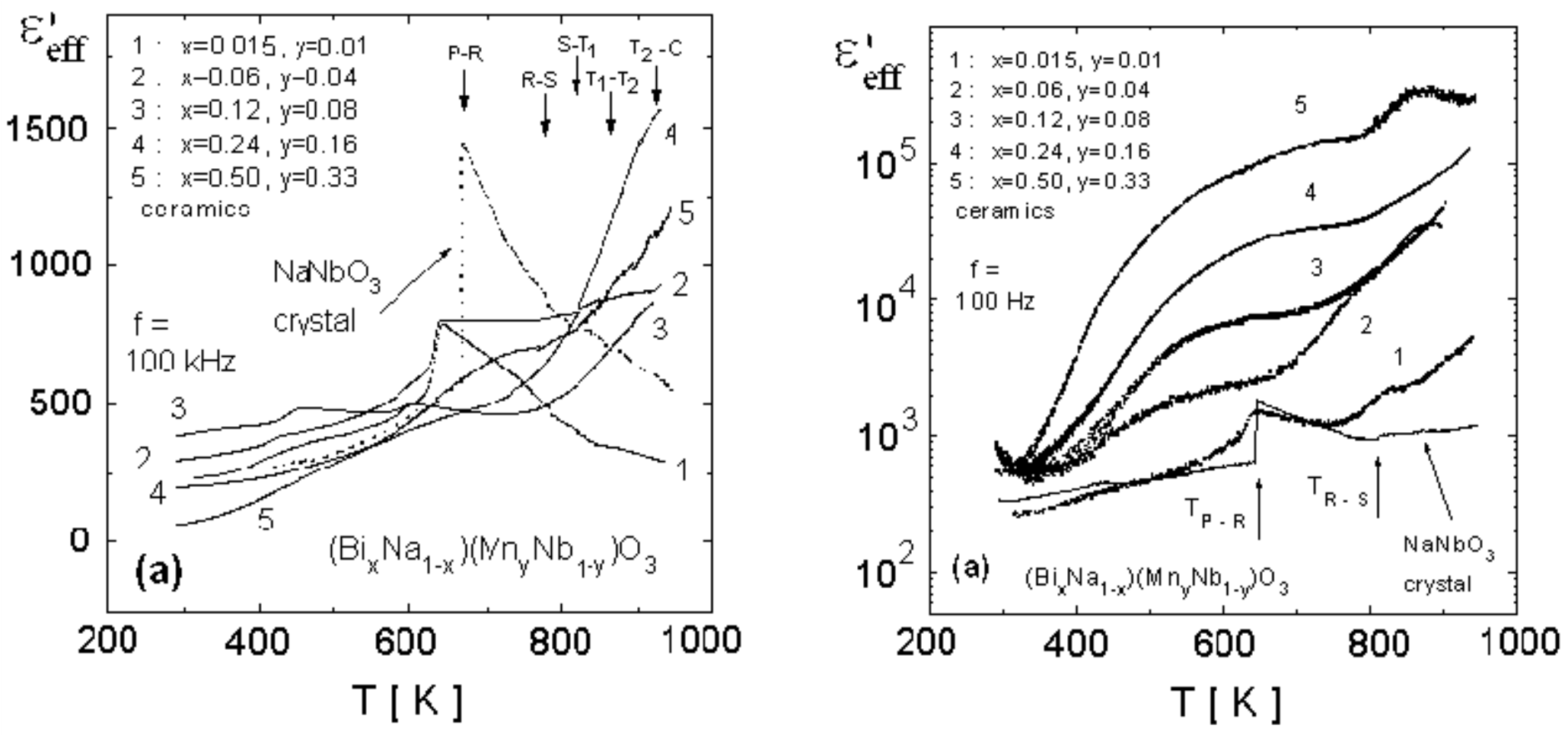}
}
\caption{Electric effective permittivity $\varepsilon '_\mathrm{eff}$ measured on heating at $f = 100$~kHz (a) and at $f = 100$~Hz (b) for (Bi$_{x}$Na$_{1-x}$)(Mn$_{y}$Nb$_{1-y}$)O$_{3}$ ceramics. The $\varepsilon '_\mathrm{eff}(T)$ obtained for a non-doped NaNbO$_{3}$ single crystal is plotted for comparison. The arrows point the temperatures of structural phase transitions occurring in pure sodium niobate between AFE and PE phases. The content of the Bi and Mn ions, $x$ and $y$, in the ceramics is noted in the diagram.}
\label{fig4}
\end{figure}

A marked dispersion in the electric properties of the (Bi$_{x}$Na$_{1-x}$)(Mn$_{y}$Nb$_{1-y}$)O$_{3}$ ceramics series is visible in the real part of the effective electric permittivity $\varepsilon_\mathrm{eff}(f,T)$  diagram [figure~\ref{fig4}~(a) and (b)].
Moderate values measured at $f = 100$~kHz were obtained for the electric permittivity i.e., $\varepsilon_\mathrm{eff}(T,100~\text{kHz}) < 2000$ within the range $300\div 900$~K, even for the samples with high Bi-Mn content.
The Bi and Mn co-doping caused a steep increase of the  $\varepsilon_\mathrm{eff}(T,100~\text{kHz})$ value at a high temperature range. This effect corresponded to the increase in the loss factor $\tan\delta$ with temperature and it was ascribed to a thermally activated effect.
The electric permittivity  $\varepsilon_\mathrm{eff}(T)$ of the lightly doped ceramics ($x = 0.015$, $y = 0.01$), measured at $f = 100$~Hz, exhibited the values close to the values measured for the crystal  $\varepsilon_\mathrm{eff}(T,100~\text{Hz}) < 10^{3}$.
\linebreak However, it markedly increased with temperature and the Mn ion content, reaching the high value $\simeq10^{5}$ for the (Bi$_{0.24}$Na$_{0.76}$)(Mn$_{0.16}$Nb$_{0.84}$)O$_{3}$ and (Bi$_{0.50}$Na$_{0.50}$)(Mn$_{0.33}$Nb$_{0.67}$)O$_{3}$ compounds at high temperature [figure~\ref{fig4}~(b)].

Hence, the addition of Bi and Mn ions to the sodium niobate host induced the high value effective permittivity when measured at low frequency and at high temperature.

\subsection{Capacitance dependence on geometrical factors}

%

\begin{figure}[!b]
\centerline{
\includegraphics[width=0.8\textwidth]{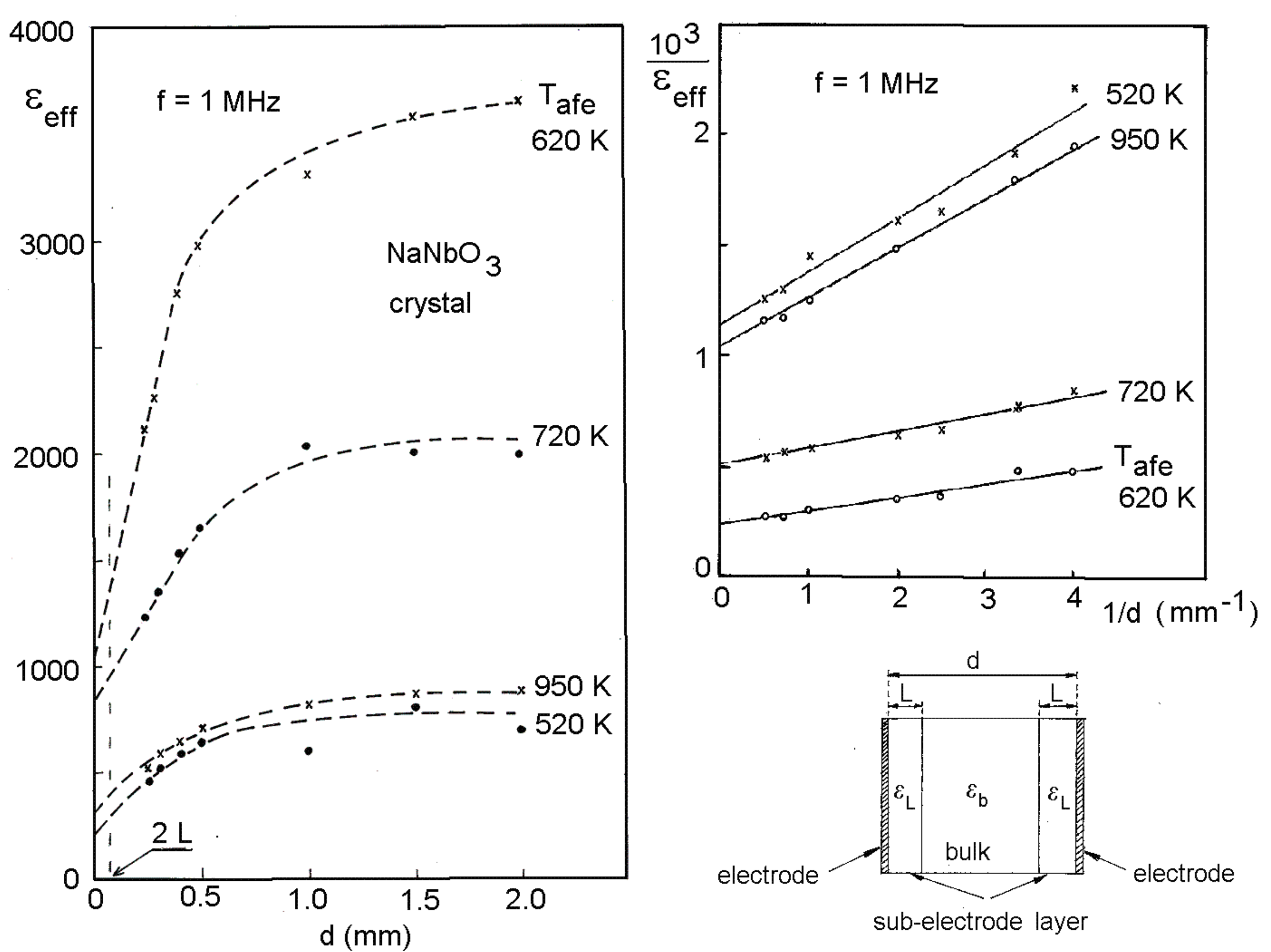}
}
\caption{
(a) The electric permittivity $\varepsilon '_\mathrm{eff}$ dependence on the sample thickness $d$ obtained for the NaNbO$_{3}$ crystals at several temperatures at $f = 1$~MHz.  $T_\textrm{afe} \approx 620$~K denotes an antiferroelectric phase transition.
(b) The reciprocal effective electric permittivity vs. reciprocal thickness of the samples.
(c) The scheme of a non-homogeneous condenser where the sub-electrode layer with thickness $L$ is assumed.}
\label{fig5}
\end{figure}

The electrostatic and the chemical capacitance show different dependences on the geometrical factors. The electrostatic capacitance is proportional to the ratio $A/d$ [equation~(\ref{e1})], while the chemical capacitance is proportional to the volume $V = Ad$ of the sample [equation~(\ref{e5})].
In case of sodium niobate, the contribution $\varepsilon _\textrm{bulk}$ from the antiferroelectric bulk can be discerned from the contribution  $\varepsilon _{L}$ of the space charge formed at the sub-electrode layers. Since the Curie-Weiss dependence is valid for an antiferroelectric phase transition, the effective electric permittivity is [figure~\ref{fig5}~(a)]
\begin{equation}
\varepsilon _\mathrm{eff} = \varepsilon _\textrm{bulk} + \varepsilon _{L} = \frac{C}{ T - T_\textrm{afe}} + \varepsilon _{L}\, .
\label{e13}
 \end{equation}								
When a non-homogeneous multi-layer model of a condenser is assumed \cite{3-Mitsui1976, 18-Suchanicz1996}, the thickness $L$ of sub-electrode layers can be estimated from the formula
\begin{equation}
\varepsilon_\mathrm{eff}^{-1} = \varepsilon_\textrm{bulk}^{-1} \left(1-\frac{2L}{d}\right)  + \varepsilon _{L}^{-1} \frac{2L}{d}   \approx \varepsilon _\mathrm{B}^{-1}   + \varepsilon _{L}^{-1}  \frac{2L}{d} \, ,
\label{e14}
 \end{equation}
where $d$ is the thickness of the sample, $L$ is the thickness of the sub-electrode layer [figure~\ref{fig5}~(c)]. The thickness $L = (1/2) \varepsilon _{L} \tan\alpha$ of the sub-electrode layer in NaNbO$_3$ crystal was estimated as $L = 40\pm10$~\SI{}{\micro\metre} (numerical fit for plots in figure~\ref{fig5}~(b), correlation coefficient $0.964\div0.990$). The $\varepsilon _L$ contribution varied from $\simeq300$ to $\simeq1000$. Therefore, the crystal lattice contribution dominated in case of NaNbO$_3$ crystals, when measured at a frequency $f = 1$~MHz.

\begin{figure}[!t]
\centerline{
\includegraphics[width=0.9\textwidth]{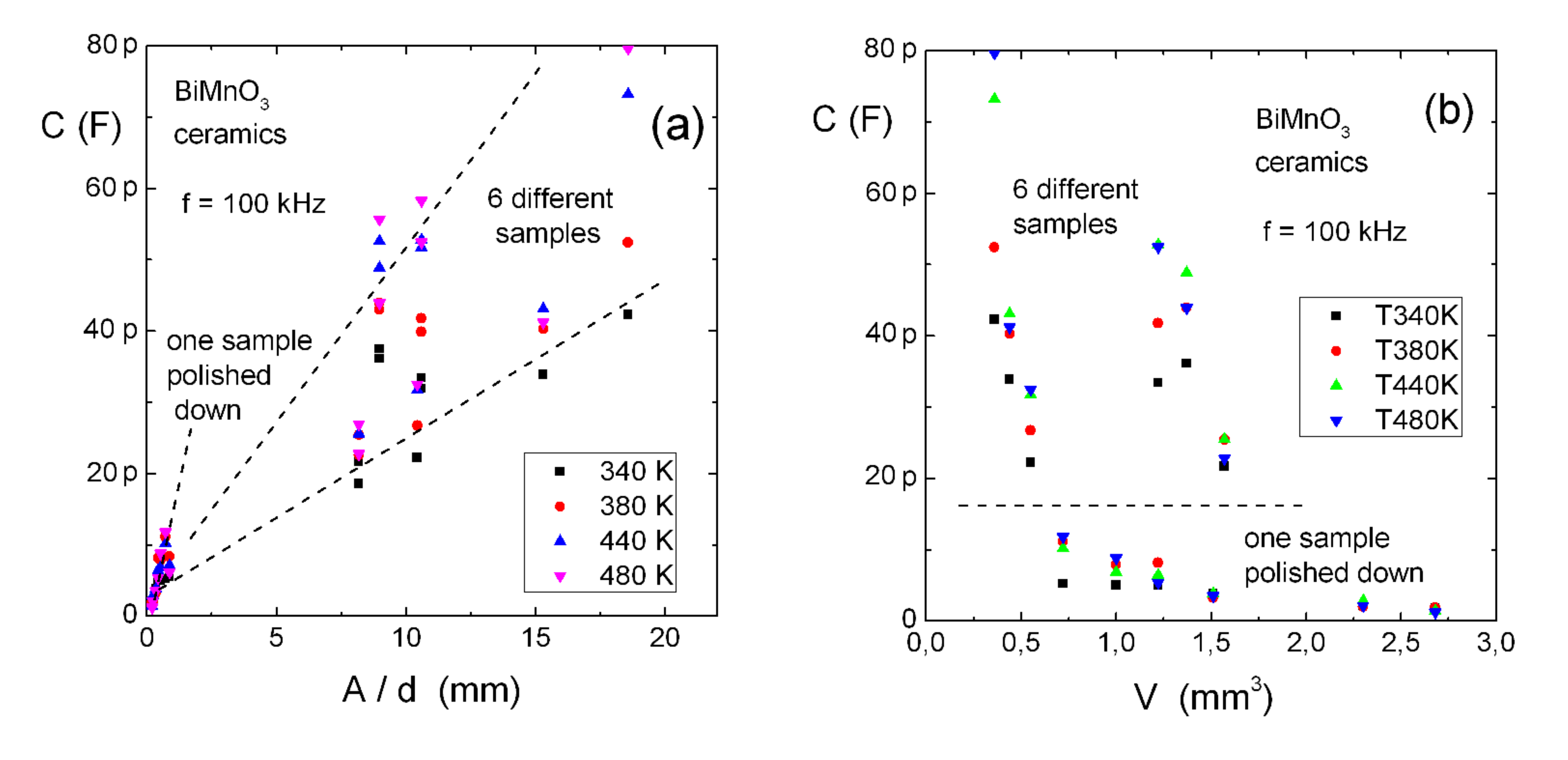}
}
\vspace{-2mm}
\caption{(Color online) (a) Geometrical dependencies  of the capacitance $C$ obtained for the BiMnO$_{3}$ ceramics measured at $f = 100$~kHz at several temperatures. (a) The dependence of $C$ on the ratio $A/d$. (b) The dependence of $C$ on the sample volume $V = Ad$.}
\label{fig6}
\end{figure}

\begin{figure}[!b]
\centerline{
\includegraphics[width=0.9\textwidth]{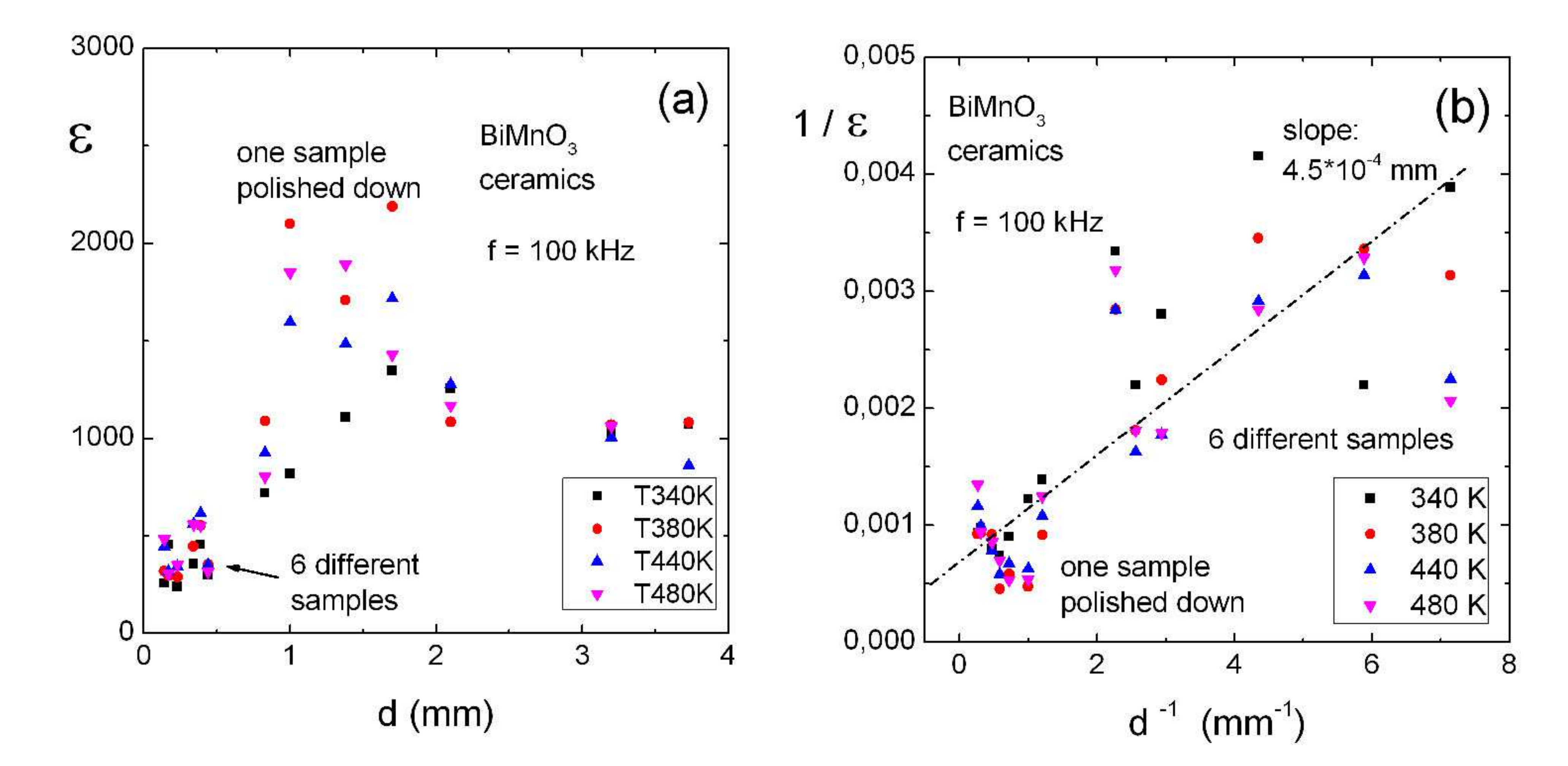}
}
\vspace{-2mm}
\caption{(Color online) (a) The effective electric permittivity $\varepsilon '_\mathrm{eff}$ dependence on the sample thickness $d$ obtained for the BiMnO$_{3}$ ceramics at several temperatures at $f = 100$~kHz. (b) The reciprocal effective electric permittivity vs. reciprocal thickness of the samples.}
\label{fig7}
\end{figure}

The dependence of capacitance on the geometrical factors, obtained for the BiMnO$_3$ ceramics, is shown in figure~\ref{fig6}. It turned out that the capacitance measured at $f = 100$~kHz increased with the $A/d$ ratio [figure~\ref{fig6}~(a)]. The $C$ dependence on the  volume of the samples was not clear.
However, a decreasing tendency can be deduced from the data obtained for the sample whose thickness was decreased by polishing step by step when $V \in (0.59\div2.68)$~mm$^{3}$. Similar results were obtained at $f = 1$~kHz. Hence, the model of electrostatic capacitance [see equations (\ref{e3}), (\ref{e13}), and (\ref{e14})] is consistent with the data presented in figure~\ref{fig6}. It should be noted that they were obtained in the $300\div500$~K range, where the dependence of the measured capacitance on temperature is weak (compare figure~\ref{fig2}).

In case of bismuth manganite, the thickness $L$ of sub-electrode layers was estimated using the data presented in figure~\ref{fig7} fitted with the equation (\ref{e14}). Assuming the value $\varepsilon _{L}  \sim200$, extrapolated for thin samples, and the slope $\tan\alpha = 4.5\cdot 10^{-4}$~mm, the sub-electrode thickness $L = (1/2)\varepsilon_{L} \tan\alpha = 50\pm30$~\SI{}{\micro\metre} was estimated.

\section{Discussion}

The studied ceramics BiMnO$_3$, (BiNa)(MnNb)O$_3$, and (BiPb)(MnNb)O$_3$ show thermally activated capacitance in a high temperature range. Such a behaviour corresponds to the semiconductor-type features of the electric conductivity. When the contribution of  thermally activated charges to the measured capacitance is assumed, the chemical capacitance model can be applied. In accord with this model, the exponential dependence on temperature
$C_{\mu}^{(\textrm{cb})} = e^{2} n_\mathrm{C}/k_\mathrm{B}T = (e^{2}N_\mathrm{C}/k_\mathrm{B}T)\exp[(\Delta E)/k_\mathrm{B}T]$ [equation (\ref{e12})]  describes the capacitance temperature dependence.
One can notice that the estimated values of parameter $\Delta E \in 0.06\div2.52$~eV
correspond to the activation energy $E_\textrm{a}$ obtained from the electric conductivity temperature dependence since the $E_\textrm{a}$ value varies within $0.2\div1.1$~eV range.  Moreover, in case of the BiMnO$_3$, the values of the $\Delta E$ can be compared to the energy gap $E_\textrm{g}$ obtained from optical absorption measurements 1.1~eV and 1.6~eV  \cite{19-Lee2010} as well as to the result obtained from ab initio calculations of the electronic structure where $E_\mathrm{g} = 0.33$~eV \cite{20-McLeod2010}.

In case of the perovskite structure, the crystal lattice component is usually on the $\varepsilon '_\mathrm{lattice} \sim100$ and higher values $\simeq 10^{3}\div10^{4}$ occur in the vicinity of structural phase transitions. From this point of view, the common origin for higher values of conduction and the capacitance temperature dependence, apart from the narrow energy gap, would be the occurrence of  defects. Such a deduction is consistent with the results of former studies which proved the occurrence of a local structural disorder and chemical non-homogeneity in these materials by XRD and EPMA tests. The estimated density of states in the vicinity of the Fermi energy was  $\sim 10^{18}\div10^{20}$~eV$^{-1}$cm$^{-3}$ \cite{4-Molak2005, 5-Molak2006, 6-Molak2005, 14-Molak2008,15-Spolnik2005}.
Hence, the high value of the measured capacitance and thus the estimated effective electric permittivity are consistent with the high concentration of defects in the samples studied, since $C_\textrm{chem} = [ (e z)^{2} /k_\mathrm{B} T] c A d $
[equation (\ref{e5})].

On the other hand, the chemical capacitance and the electrostatic capacitance show different dependences on the geometrical factors. Therefore, the test was conducted for the end members of the studied series of compounds, i.e., for NaNbO$_3$ and BiMnO$_3$.

In case of the stoichiometric sodium niobate crystals, the chemical capacitance contribution was not expected. The dielectric permittivity did not increase with temperature increase. However, a low concentration of the space charge was determined $\sim10^{17}$~cm$^{-3}$.
Hence, the model of electrostatic, non-homogeneous, multi-layer condenser was used to estimate the participation of the space charge in the effective permittivity. The thickness of sub-electrode layers was of the order of several tenth of~\SI{}{\micro\metre}.

In case of sodium niobate and bismuth manganite, one can expect a correlation of the capacitance with the volume of the samples, in accord with the chemical capacitance model.  However, it turned out that the capacitance was not proportional to the volume of the samples when they were measured and analysed in the temperature range $300\div500$~K.  Instead, it was proportional to the $A/d$ ratio which indicated that the electrostatic, non-homogeneous model is appropriate for bismuth manganite. The estimated thickness of  sub-electrode layers was also of the order of several tenth of~\SI{}{\micro\metre}.

Hence, in case of BiMnO$_3$, contradictory results were obtained from the analysis of the geometrical factor dependence and from the high temperature dependence of the capacitance. One can notice that the geometrical factors analysis was carried out in the $300\div500$~K range where the capacitance dependence on temperature was weak.

Nevertheless, the chemical capacitance model should be used for high temperature ranges  for  (Bi$_{0.5}$Na$_{0.5}$)(Mn$_{0.5}$Nb$_{0.5}$)O$_3$, (Na$_{0.5}$Pb$_{0.5}$)(Mn$_{0.5}$Nb$_{0.5}$)O$_3$, BiMnO$_3$ compounds (see figure~\ref{fig1} and \ref{fig2}) where a steep increase in $C(T)$ took place. The thermal generation of the charge carriers, which contribute to the capacitance and induce its high value, is likely to occur only in a part of the volume of the samples. Therefore, it seems that further studies conducted at $T > 600$~K would be quite useful in order to discern the ambiguities.

\ukrainianpart

\title{Хімічна ємність, запропонована для керамік на основі манганіту}

\author{А.~Моляк}
\address{Інститут фізики, Сілезький університет, Катовіце, Польща}

 \makeukrtitle
\begin{abstract}

Виміряне значення ефективної електричної проникності
$\varepsilon_\mathrm{eff}$ декількох сполук, а саме,
(BiNa)(MnNb)O$_3$, (BiPb)(MnNb)O$_3$, і BiMnO$_3$ збільшилося із
значення $\approx 10\div100$ в низькотемпературній області
($100\div300$~K) до високого значення, досягаючи значення $10^5$ у
високотемпературній області, а саме, $500\div800$~K. Такі риси є
виявом термічно активованих носіїв просторового заряду, які впливають
на вимірювану ємність. Виміряне високе значення ефективної
проникності декількох магнітних сполук може бути приписане
хімічній ємності
 $C_{\mu} = e^2 \partial N_{i}/\partial\mu_i $, вираженій в термінах
 хімічного потенціалу  $\mu$. Хімічна ємність
$C_{\mu}^{(cb)} = e^2 n_\mathrm{C}/k_\mathrm{B}T$ залежить від
температури, при якій розглядаються електрони провідності з густиною
$ n_\mathrm{C} = N_\mathrm{C} \exp\left(\mu_{n}-
E_\mathrm{C}\right)/k_\mathrm{B}T$. Експериментальні результати,
отримані для сполук манганіту у високотемпературній області,
обговорюються в рамках моделі хімічної ємності. Проте, виміряна
ємнісна залежність від геометричних факторів, що аналізується для
BiMnO$_3$, вказує, що неоднорідна електростатична ємнісна модель є
справедливою в області $300\div500$~K.
\keywords хімічна ємність, електрична проникність, перовскіти,
дефекти
\end{abstract}

\clearpage

\begin{thebibliography}{99}
\bibitem{1-Sherill2011}Sherill~S.A.,  Banerjee~P., Rubloff~G.W.,  Lee~S.B.,  Phys. Chem. Chem. Phys., 2011, \textbf{13}, 20714; \\ \doi{10.1039/C1CP22659B}.

\bibitem{2-Molak2009} Molak~A., Pawelczyk~M., Kubacki~J.,  Szot~K., Phase Transitions, 2009, \textbf{82}, 662;  \doi{ 10.1080/01411590903341155}.

\bibitem{3-Mitsui1976} Mitsui~T.,  Tatsuzaki~I.,  Nakamura~E., An introduction to the physics of ferroelectrics, Gordon and Breach, Science Publ., New York, 1976.

\bibitem{4-Molak2005}Molak~A., Paluch~M.,  Pawlus~S.,  Klimontko~J.,  Ujma~Z.,  Gruszka~I.,  J. Phys. D: Appl. Phys., 2005, \textbf{38}, 1450; \doi{10.1088/0022-3727/38/9/019}.

\bibitem{5-Molak2006}Molak~A., Paluch~M.,  Pawlus~S., Ujma~Z., Pawelczyk~M.,  Gruszka~I.,  Phase Transitions, 2006, \textbf{79}  447; \\ \doi{10.1080/01411590600892336}.

\bibitem{6-Molak2005} Molak~A., Ksepko E., Gruszka~I.,  Ratuszna~A.,  Paluch M.,  Ujma Z., Solid State Ionics, 2005, \textbf{176}, 1439; \\ \doi{10.1016/j.ssi.2005.03.013}.

\bibitem{7-Jamnik2001} Jamnik~J., Meier~J.,  Phys. Chem. Chem. Phys., 2001, \textbf{3},  1668; \doi{10.1039/b100180i}.

\bibitem{8-Bisquert2003} Bisquert J., Phys. Chem. Chem. Phys., 2003, \textbf{5}, 5360; \doi{10.1039/b310907k}.

\bibitem{9-Fabregat2011} Fabregat-Santiago~F.,  Garcia-Belmonte G.,  Mora-Sero I.,  Bisquert~J.,  Phys. Chem. Chem. Phys., 2011, \textbf{13}, 9083; \doi{10.1039/c0cp02249g}.

\bibitem{10-Baumann2006} Baumann~F.S., Fleig~J.,  Habermeier~H.U.,  Maier~J.,  Solid State Ionics, 2006, \textbf{177}, 1071; \doi{10.1016/j.ssi.2006.07.057}.

\bibitem{11-Bisquert2008} Bisquert J., Phys. Rev. B, 2008, \textbf{77}, 235203; \doi{10.1103/PhysRevB.77.235203}.

\bibitem{12-Bisquert2008}Bisquert~J.,  Phys. Chem. Chem. Phys., 2008,  \textbf{10}, 49; \doi{10.1039/b709316k}.

\bibitem{13-Badurski1979} Badurski~M.,  Handerek~J.,  Szatanek~J.,  Szot~K., Acta Phys. Pol. A, 1979, \textbf{55},  855.

\bibitem{14-Molak2008} Molak~A., Pawelczyk~M., Ferroelectrics, 2008, \textbf{367}, 179; \doi{10.1080/00150190802375417}.

\bibitem{15-Spolnik2005} Spolnik~Z.,  Osan~J.,  Klepka~M.,  Lawniczak-Jablonska~K.,  van Grieken~R.,  Molak~A.,  Potgieter~J.H., Spectrochimica Acta B, 2005, \textbf{60}, 525; \doi{10.1016/j.sab.2005.03.013}.

\bibitem{16-Belik2007} Belik~A.A.,  Iikubo~S., Yokosawa~T.,  Kodama~K.,  Igawa~N.,  Shamoto~Sh.,  Azuma~M.,  Takano~M.,  Kimoto~K., Matsui~Y.,  Takayama-Muromachi~E., J. Am. Chem. Soc., 2007, \textbf{129}, 971; \doi{10.1021/ja0664032}.

\bibitem{17-Kimura2003} Kimura~T.,  Kawamoto S.,  Yamada I.,  Azuma M.,  Takano M., Tokura~Y.,  Phys. Rev. B, 2003, \textbf{67}, 180401; \\ \doi{10.1103/PhysRevB.67.180401}.

\bibitem{18-Suchanicz1996} Suchanicz~J.,  Molak~A.,  Kus~C.,  Ferroelectrics, 1996, \textbf{177}, 201; \doi{10.1080/00150199608223629}.

\bibitem{19-Lee2010} Lee~J.H.,  Ke X.,  Misra~R., Ihlefeld~J.H.,  Xu~X.S.,  Mei~Z.G.,  Heeg~T.,  Roeckerath~M.,  Schubert~J.,  Liu~Z.K.,  Musfeld~J.L.,  Schiffer~P.,  Schlom~D.G.,  Appl. Phys. Lett., 2010, \textbf{96},  262905; \doi{10.1063/1.3457786}.

\bibitem{20-McLeod2010} McLeod~J.A.,  Pchelkina Z.V., Finkelstein~L.D.,  Kurmaev~E.Z.,  Wilks~R.G.,  Moewes~A., Solovyev~I.V.,  Belik~A.A.,  Takayama-Muromachi~E., Phys. Rev. B, 2010, \textbf{81}, 144103; \doi{10.1103/PhysRevB.81.144103}.

\end{thebibliography}
 \end{document}